\documentstyle[twoside,fleqn,npb,epsfig]{article}
\newcommand{\AmS}{{\protect\the\textfont2
  A\kern-.1667em\lower.5ex\hbox{M}\kern-.125emS}}

\newcommand{\GeV}{{\, \rm GeV}}

\title{Direct determination of the gluon density in the proton
       from jet cross sections in deep-inelastic scattering}

\author{M. Wobisch\address{III. Physikalisches Institut, 
 RWTH Aachen, D-52056 Aachen, Germany} for the H1 collaboration}

\begin{document}
\pagestyle{empty}

\begin{abstract}
A QCD analysis of jet production in deep-inelastic scattering
is presented.
The studies are based on recent measurements of jet cross sections 
in the H1 experiment at HERA which are directly sensitive to the gluon 
density in the proton.
The inclusion of H1 data on the inclusive deep-inelastic scattering 
cross section enables us to perform a consistent, simultaneous determination of 
the quark and gluon densities in the proton.
The gluon density has been determined
at fairly large values of $10^{-2} < x < 10^{-1}$ at a scale
of the order of the transverse jet energies $\mu^2_f = 200\GeV^2$.
The result is consistent with indirect determinations
from HERA structure function data 
and extends their range of sensitivity to larger $x$-values.
\end{abstract}

\pagenumbering{roman}
\hfill \parbox[t]{4cm}{\large
PITHA 99/17 \\
hep-ph/9906399
}

\vskip10mm

\begin{center}
{\bf \huge
Direct determination of the gluon density in the proton \\ \vskip2mm
       from jet cross sections in deep-inelastic scattering}

\vskip20mm

{\LARGE M. Wobisch \\ \vskip4mm
 \Large III. Physikalisches Institut, RWTH Aachen,\\ \vskip2mm
        D-52056 Aachen, Germany}

\end{center}

\vskip20mm

\begin{center}
\large \bf Abstract \\ \vskip1mm
\parbox[t]{14.5cm}{\large
A QCD analysis of jet production in deep-inelastic scattering
is presented.
The studies are based on recent measurements of jet cross sections 
in the H1 experiment at HERA which are directly sensitive to the gluon 
density in the proton.
The inclusion of H1 data on the inclusive deep-inelastic scattering 
cross section enables us to perform a consistent, simultaneous determination of 
the quark and gluon densities in the proton.
The gluon density has been determined
at fairly large values of $10^{-2} < x < 10^{-1}$ at a scale
of the order of the transverse jet energies $\mu^2_f = 200\GeV^2$.
The result is consistent with indirect determinations
from HERA structure function data
and extends their range of sensitivity to larger $x$-values.}
\end{center}

\vskip31mm
\noindent
Talk given on behalf of the H1 collaboration at the 7th
International Workshop on Deep-Inelastic Scattering
and QCD (DIS99), Zeuthen, April 1999.

\newpage
\pagenumbering{arabic}

\maketitle

\pagestyle{plain}
\pagenumbering{arabic}

\section{INTRODUCTION}
The present knowledge on the gluon density in the proton 
basically comes from deep-inelastic scattering (DIS) 
structure function data (i.e. from inclusive DIS cross sections).
These are, however, only indirectly
sensitive to the gluon density, which enters the cross section
formulae only via the next-to-leading order (NLO) corrections
in the boson-gluon fusion process.

A process that is {\em directly} 
sensitive to the gluon density in the proton is the production of jets
at (moderately) high transverse energies ($E_T$) in the Breit frame.
At leading order, high $E_T$ jet cross sections are described
by QCD-Compton and by boson-gluon fusion processes, the latter
being dominant over large regions of phase space.
A QCD analysis of jet cross sections may therefore lead to
a direct determination of the gluon density, independent of 
assumptions needed in the indirect determinations from structure 
function data.

\section{EXPERIMENTAL RESULTS}

The dijet and the inclusive jet cross sections presented here 
have been measured with the H1 detector at HERA, based on data taken 
in the years 1994-97 corresponding to an integrated luminosity
${\cal L}_{\rm int} \simeq 36\,{\rm pb}^{-1}$.
The kinematic range extends from moderate to large 
momentum transfers $10 < Q^2 < 5000\GeV^2$ for $0.2 < y < 0.6$.
Jets are defined by the inclusive $k_\perp$ algorithm \cite{inclkt}
which is applied to the final state particles in the Breit frame.
Only jets in the central region of the detector acceptance with 
pseudorapidities $-1 < \eta_{\rm jet, lab} < 2.5$
and transverse jet energies $E_{T {\rm Breit}} > 5\GeV$ are considered.

\begin{figure}[t]
\epsfig{file=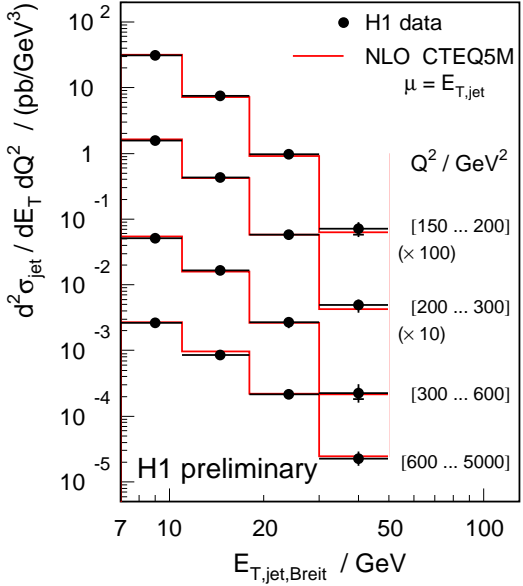,width=7.2cm}
\vskip-8mm
\caption{The inclusive jet cross section as a function of the 
transverse jet energy $E_T$ in the Breit frame in different regions 
of $Q^2$.} 
\label{fig:incljetet}
\end{figure}

\begin{figure}[t]
\epsfig{file=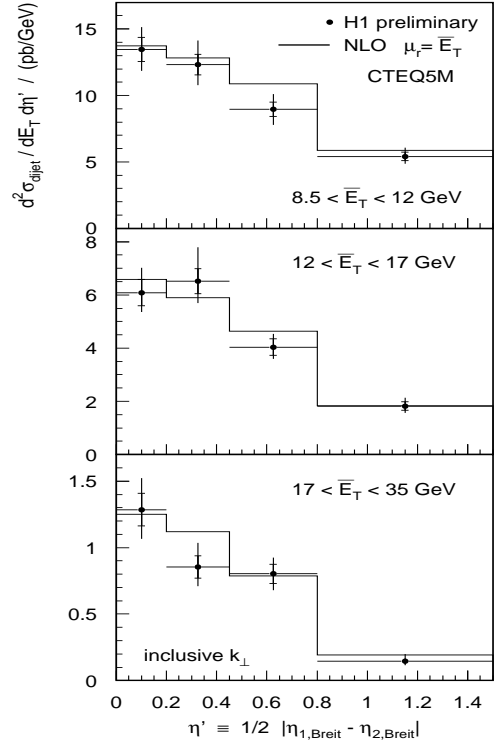,width=6.8cm,height=10.1cm}
\vskip-8mm
\caption{The dijet cross section in different regions of the average
transverse jet energy $\overline{E}_T$ in the Breit frame
as a function of the jet pseudorapidity $\eta'$ (reconstructed
from the pseudorapidity difference of the jets in the Breit frame 
$\eta' = 1/2 \, | \eta_1 - \eta_2 | $).
The measurement is performed at $200 < Q^2 < 5000\GeV^2$.}
\label{fig:dijetetap}
\end{figure}

\begin{figure}[t]
\epsfig{file=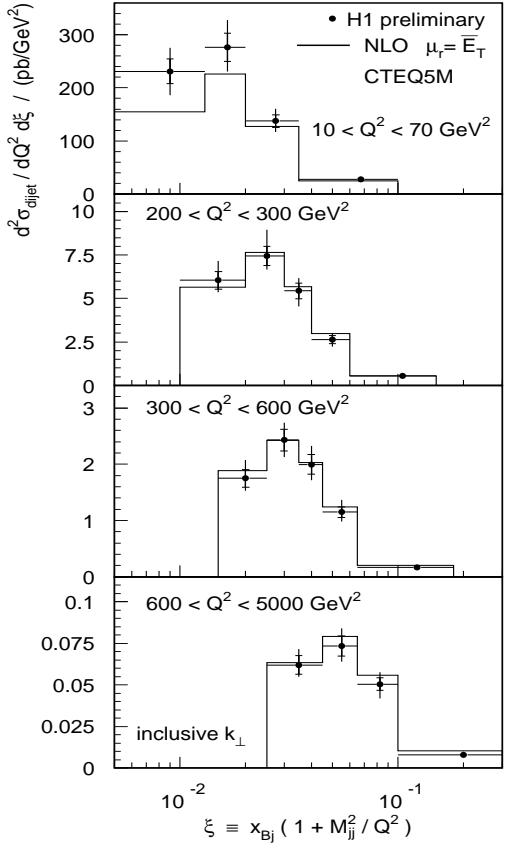,width=7cm,height=11.4cm}
\vskip-8mm
\caption{The dijet cross section in different regions
of $Q^2$ as a function of the fractional parton momentum $\xi$
reconstructed from the invariant mass of the dijet system.}
\label{fig:dijetxi}
\end{figure}

The inclusive jet cross section 
${\rm d}^2 \sigma_{\rm jet} / {\rm d}E_T {\rm d}Q^2$
has been measured for $7 < E_{T \rm jet, Breit} < 50\GeV$.
For the dijet cross section events with two or more jets 
have been selected where the two jets with the highest $E_T$ 
fulfill $\sum E_T > 17\GeV$. 
Double differential distributions have been obtained for a large 
set of variables \cite{ichep520}.
Here we show the dependence of the inclusive jet cross section
on both hardness scales $E_T$, $Q^2$ (Fig.\ \ref{fig:incljetet}),
and the dependence of the dijet cross section on the jet
pseudorapidity $\eta'$ and the 
reconstructed fractional parton momentum $\xi$
(Figs.\ \ref{fig:dijetetap},\ref{fig:dijetxi}).
The basic observations are:
towards larger $Q^2$ a harder $E_T$ spectrum is seen;
large jet pseudorapidities $\eta'$ are suppressed at higher $E_T$;
the jet cross sections are sensitive to  fractional parton momenta 
$0.005 < \xi < 0.3$.

The perturbative calculations give a very good description of 
these and other distributions \cite{ichep520}, except at small 
values of $Q^2 < 100\GeV^2$ where NLO corrections are very large
(k-factors are between 1.6 and 2) and sizeable contributions 
from higher orders are expected.

\section{QCD ANALYSIS}
We have seen that the jet data are nicely described by 
perturbative calculations at NLO when parton densities and 
$\alpha_s(M_Z)$ values are taken from global fits.
We therefore conclude that perturbative QCD at 
next-to-leading order $\alpha_s$ is able to describe 
jet production processes in DIS, at least in the kinematic
region under investigation: at fairly large transverse jet energies 
$E_T$ in the Breit frame and momentum transfers $Q^2$ not 
too small.
The influence of non-perturbative contributions has been 
investigated \cite{hadcor} and is found to be small (below 7\% for 
the dijet cross sections) with negligible model dependence. 

It is now straightforward to make a QCD analysis of these data
for a determination of the free parameters of the theory: 
$\alpha_s$, the gluon density and the quark densities in the proton.
The present data are, however, not able to constrain these 
simultaneously.
We therefore decided to take the value of $\alpha_s(M_Z)$ 
(as e.g.\  measured by the LEP experiments independently 
of the proton structure)
as external input, and perform a simultaneous fit
of the gluon and quark densities in the proton.
For this purpose we include H1 data on the inclusive DIS 
cross section \cite{ichep533} in the QCD analysis
at $Q^2$ values which are of the order of the transverse 
jet energies $E^2_T$ in the dijet cross sections.

In this kinematic region ($200 < Q^2 < 650\GeV^2$) 
the inclusive DIS cross sections give very strong constraints 
on the quark densities while they depend on the 
gluon density only weakly.
In the combined fit of both datasets the inclusive DIS cross sections
therefore determine the quarks while the dijet cross sections 
determine the gluon density. 

\begin{figure}[t]
\epsfig{file=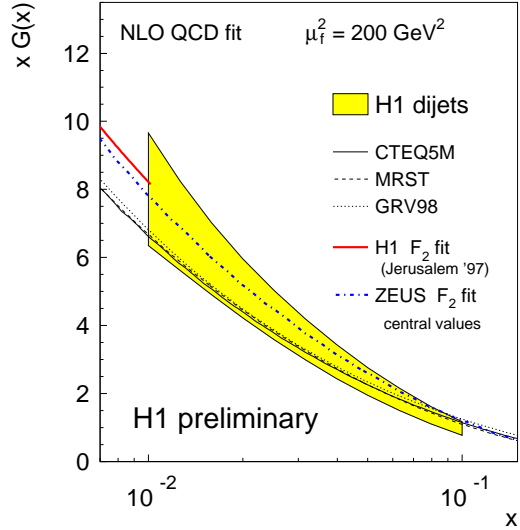,width=7cm}
\vskip-11mm
\caption{The gluon density in the proton in the $\overline{\rm MS}$-scheme
determined in a NLO QCD fit to dijet cross sections at a factorization
scale $\mu^2_f = 200\GeV^2$ for a value of $\alpha_s(M_Z)=0.119\pm0.005$.
The error band is compared to results from global fits and to the results
of QCD fits to HERA structure function data.
}
\label{fig:gluon}
\end{figure}

The dijet data used in the fit are restricted to $Q^2 > 200\GeV^2$ 
where NLO corrections and hadronization corrections are small.
The fit has been performed to the 
double differential cross sections
${\rm d}^2 \sigma_{\rm dijet} / {\rm d}Q^2 {\rm d}\xi$,
${\rm d}^2 \sigma_{\rm dijet} / {\rm d}Q^2 {\rm d}x_{\rm Bj}$
and 
${\rm d}^2 \sigma_{\rm incl.} / {\rm d}Q^2 {\rm d}x_{\rm Bj}$
which are most sensitive to the $x$-dependence of the 
parton distributions.
The gluon and the quark densities have been fitted at a factorization
scale $\mu^2_f = 200\GeV^2$ which is of the size of the hard scales
for both the dijet ($\mu^2_f \simeq E^2_T$) and the inclusive DIS 
cross sections ($\mu^2_f \simeq Q^2$).
More details on the fitting procedure can be found in
\cite{ichep520}.

The resulting gluon density is displayed in Fig.~\ref{fig:gluon}
at the scale $\mu^2_f = 200\GeV^2$ in the range $0.01 < x < 0.1$.
The error band includes experimental and theoretical
uncertainties as well as the uncertainty of $\alpha_s(M_Z)$.
Within these uncertainties this direct determination 
is consistent to the results from global fits and to indirect 
determinations from HERA structure function data 
and extends their range of sensitivity to larger $x$-values.

\end{document}